\documentclass[extra,referee]{gji}
\usepackage{graphicx}
\usepackage{amsmath}

\begin{document}
\title[Modelling of SP dependence on water content in sand]{A 1{D} modelling of streaming potential dependence on water content during drainage experiment in sand}

\author[V. All\`egre, F. Lehmann, P. Ackerer, L. Jouniaux and P. Sailhac]
   {V. All\`egre$^1$, F. Lehmann$^2$, P. Ackerer$^2$, L. Jouniaux$^1$ and P. Sailhac$^1$\\
  $^{1}$Institut de Physique du Globe de Strasbourg, UdS/CNRS UMR-7516 \\Universit\'e de Strasbourg,
5 rue Ren\'e Descartes, 67084, Strasbourg, FRANCE\\
  $^{2}$Laboratoire d'Hydrologie et de G\'eochimie de Strasbourg, UdS/CNRS UMR-7517 \\Universit\'e de Strasbourg,
1 rue Blessig, 67000, Strasbourg, FRANCE}

\date{}

\maketitle
\begin{keywords}
electrokinetics, streaming potential, self potential, water saturation, unsaturated flow, water content, finite element method, drainage experiment
\end{keywords}
\begin{abstract}
\noindent
The understanding of electrokinetics for unsaturated conditions is crucial for numerous of geophysical data interpretation. Nevertheless, the behaviour of the streaming potential coefficient $C$ as a function of the water saturation $S_w$ is still discussed. We propose here to model both the Richards equation for hydrodynamics and the Poisson's equation for electrical potential for unsaturated conditions using 1D finite element method. The equations are first presented and the numerical scheme is then detailed for the Poisson's equation.
Then, computed streaming potentials (SP) are compared to recently published SP measurements carried out during drainage experiment in a sand column. We show that the apparent measurement of $\Delta V / \Delta P$ for the dipoles can provide the streaming potential coefficient in these conditions. Two tests have been performed using existing models for the SP coefficient and a third one using a new relation. The results show that existing models of unsaturated SP coefficients $C(S_w)$ provide poor results in terms of SP magnitude and behaviour. We demonstrate that the unsaturated SP coefficient can be until one order of magnitude larger than $C_{sat}$, its value at saturation. We finally prove that the SP coefficient follows a non-monotonous behaviour with respect to water saturation.
\end{abstract}     
\section{Introduction}
\noindent
Electric and electromagnetic methods are used in a large range of geophysical applications because of their sensitivity to fluids within the crust. The electrical resistivity can be related to the permeability and to the deformation, in full-saturated or in partially-saturated conditions \citep{doussan09,hen03,joun94,joun06}.
The self-potentials have also been successfully used in the last decades.
Self-potentials have been observed to detect contaminant plumes or salted fronts through the interpretation of electrochemical effects \citep{naudet03,maineult06a,maineult06b}.
Most of the self-potential observations are interpreted through the electrokinetic effect. It has been proposed to use this electrokinetic effect for the seismic prediction \citep{fenoglio95,pozzi94}. For hydrological applications, some hydraulic properties can be inferred from the self-potential observations \citep{gibert01,sail04,glover06,glover09}.
The W-shaped anomalies classically observed on active volcanoes are
used to characterize geothermal circulations \citep{finizola02,finizola04,saracco04,mauri10,onizawa09}.
It has also been proposed to use the self-potential monitoring to detect at distance the propagation of a water front in a reservoir \citep{saunders08}.
Moreover self-potentials have been monitored during hydraulic tests in boreholes leading to some relationship with the microseismicity \citep{darnet06}, and showing a non-linear behaviour that could be related to the saturation and desaturation processes \citep{main08}.
Streaming potential (SP) results from the coupling between fluid (water) flow and electrical current, through the motion of ionic charges of water in the pore space. The distribution of ions near the matrix surface is described by the electric double layer, including the diffuse layer in which the number of counterions exceeds the number of cations adsorbed to the matrix \citep{davis78}. The streaming current is caused by the motion of ions from the diffuse layer, coming from a pressure difference. This current is then balanced by a conduction current leading to the SP.\\
In steady state flows through homogeneous media, one can define the streaming potential coefficient $C$ as the ratio between the measured electrical potential difference $\Delta V$ and the driving pore water pressure difference $\Delta P$ \citep{over52}. The streaming potential is a function of various parameters, and its dependence on water salinity \citep{ishi81,jaafar09,vinogradov10}, water electrical conductivity \citep{pride91,lorne99a}, pH \citep{ishi81,guichet06} or temperature \citep{tosha03} is still studied.\\
However, one is forced to notice a lake of data concerning the SP coefficient dependence on water content.
For shallow surface geophysical applications, including also the seismoelectric conversions \citep{dupuis07} which is studied in laboratory \citep{bordes06,bordes08}, a better understanding of electrokinetics for unsaturated conditions is needed. A way for the understanding of such a phenomena is to study the streaming potential coefficient as a function of saturation.\\
The first experimental SP coefficient measurements were performed by \citet{guichet03}. These authors measured SP during drainage experiments performed by injecting inert gas through sand, and inferred a linear relation between the relative SP coefficient $C_r$ (i.e $C$ normalized by its value at saturation $C_{sat}$) and effective water saturation $S_e$. \citet{perrier00} proposed an empirical expression to explain the dependence of $C_r$ on water content based on a relative permeability model. The implicit assumption was that the electrical currents are affected by unsaturated state in a comparable way than hydrological flow. \citet{revil07} proposed recently another formula to characterize this dependence also based on a relative permeability model. \citet{linde07}  proposed another expression to model some SP measurements performed during a drainage experiment, with similar conditions to those from \citet{allegre2010}. However, these studies do not provide a combined hydrodynamic and electrical approach to model the data, as it should be done. For reservoir applications \citet{saunders08} proposed a linear expression for $C_r$ in the case of oil imbibition. All these models predict a monotonous decrease of $C_r$ with decreasing water saturation. Recently, \citet{allegre2010} proposed original SP measurements performed during a drainage experiment and measured the first continuous recordings of the SP coefficient as a function of water saturation. They observed that the SP coefficient exhibits two different behaviours as the water saturation decreases. Values of $C_r$ first increase for decreasing saturation in the range $0.55-0.8 < S_w < 1$, and then decrease from $S_w = 0.55-0.8$ to residual water saturation. This behaviour was never reported before and called for new interpretations of electrokinetic phenomena for unsaturated conditions.\\
Streaming potentials have been successfully modelled for aquifer properties determination \citep{darn03} or for water infiltration conditions \citep{sail04}. \citet{sheffer07} proposed a 3D modelling of streaming potential at the field scale
for saturated conditions. \citet{jackson2010} used a bundle capillary model to compute the SP coefficent as a function of water-saturation. He showed that the behaviour of the SP coefficient depends on the capillary size distribution, the wetting behaviour of the capillaries, and whether we invoke the thin or thick electrical double layer assumption. Depending upon the chosen value of the saturation exponent and the irreductible water-saturation, the relative SP coefficient may increase at partial saturation, before decreasing to zero at the irreductible saturation.
\\
Finally, no previous study took into account hydrodynamics and electrical potential equations together, even in a simple geometry, to model the streaming potential coefficient for unsaturated conditions. Thus, we propose here to model SP by solving both the Richards equation for hydrodynamics and the Poisson's equation for electrical potential using 1D finite element method. Existing models which describe the behaviour of $C_r$ as a function of $S_e$ are tested and compared to a new expression inferred from SP measurements by \citet{allegre2010}. Thus, after introducing governing equations, computed SP using these models are presented and compared to measurements. The results lead to the conclusions that 1) a non-monotonous behaviour of $C_r$ is required to fit the measurements; 2) The apparent measurement of $\Delta V / \Delta P$ for the dipoles can provide the streaming potential coefficient in these conditions.
\section{Governing equations}
\subsection{Hydrodynamic equations}
\noindent
Combining the mass conservation equation to the 1D generalized Darcy's law leads to the mixed form of the Richards equation \citep{rich31}, which describes unsaturated flow in porous media,
\begin{equation}
\frac{\partial \theta (h)}{\partial t}  - \frac{\partial}{\partial z} \left[ K(h)\left(\frac{\partial h}{\partial z} - 1 \right) \right] = 0
\label{eqn:richards}
\end{equation}
where $\theta(h)$ is the volumetric water content [m$^3$.m$^{-3}$], dependent on the pressure head $h$ [m]. The parameter $K$, which is also a function of the pressure head, is the hydraulic conductivity [m.s$^{-1}$], $t$ is time [s], and $z$ is the vertical coordinate [m] taken to be positive downward. The dependence of the hydraulic conductivity and pressure head on water content is non-linear. Numerous retention and relative permeability models are able to take into account this dependence \citep{gardner58,brooks64,vanGen80}. The models which have been chosen for this work were proposed respectively by \citet{brooks64},
\begin{equation}
Se = \frac{\theta - \theta_r}{\theta_s - \theta_r}=
\left\{
\begin{array}{l}
{{\left(\dfrac{h_a}{|h|}\right)^{\lambda}}~~,\text{if}~~\dfrac{h_a}{|h|} < 1}\\
{1~~,\text{if}~~\dfrac{h_a}{|h|} > 1}
\end{array}
\right.
\label{eqn:brookscorey}
\end{equation}
and \citet{mual76},
\begin{equation}
K(S_e)  = {K_s}.{S_e}^{L+2+\frac{2}{\lambda}}
\label{eqn:mualem}
\end{equation}
with $S_e$ the effective water saturation, $\theta_s$ the water content at saturation [m$^3$.m$^{-3}$], also equal to porosity $\phi$, $\theta_r$ the residual water content [m$^3$.m$^{-3}$], and $K_s$ the hydraulic conductivity at saturation [m.s$^{-1}$]. The effective water saturation can be expressed by: $S_e = (S_w - S_w^r) / (1 - S_w^r)$, with $S_w$ ($S_w = \theta/\phi$) and $S_w^r$ ($S_w^r = \theta_r/\phi$) the water saturation and the residual water saturation respectively. The parameter $\lambda$ in equation \ref{eqn:brookscorey} is a measure of the pore size distribution and characterizes the medium granulometry. Thus, higher the value of $\lambda$ is, higher homogeneous the medium is. The second hydrodynamic parameter $h_a$ is the air entry pressure \citep{brooks64}. The last parameter $L$ takes into account the tortuosity and is chosen as $L=0.5$, which is a common value in the literature \citep{mual76}.\\
Some initial and boundary conditions are necessary to solve the Richards equation. Initial condition is a saturated sand at the hydrostatic equilibrium, so that the water pressures (expressed in cm of water) at the top and the bottom boundary of the system are respectively $h(z=0) = 0$ and $h(z=l) = l$. Considering the drainage experiment which will be presented in this work, two conditions have been used. The first one is a zero flux $q_0(t)$ at the top of the system (Neumann type) and the second one is a constant pressure head $h_l(t)$ at the bottom (Dirichlet type). These conditions are written as,
\begin{equation}
h(z=l,t) = h_l(t)\ \ {\text{and}}\ \ \left(-K(h)\frac{\partial h}{\partial z} + K\right)_{z=0} = q_0(t)
\label{eqn:boundarycond}
\end{equation}
where $z=0$ or $z=l$ and $l$ the length of the system. The Richards equation has been discretized using the Galerkin finite element method \citep{pind77}, with a fully implicit scheme in time. To take into account dependency between $h$, $\theta$ and $K$, the equation is linearized using the Newton-Raphson method. This approach has been used for decades to solve this equation, and the detailed scheme can be found for example in \citet{leh98}.
\subsection{Electrokinetic theory}
\noindent
Coupled fluxes can be described by the general equation,
\begin{equation}
\mathbf{J}_i = \sum_{j=1}^{N} {\cal{L}}_{ij} \mathbf{X}_j
\end{equation}
which link the forces $\mathbf{X}_j$ to macroscopic fluxes $\mathbf{J}_i$, through transport coupling coefficients ${\cal{L}}_{ij}$ \citep{Ons31}. The global SP field can be described as the sum of several contributions creating electrical current sources. These current sources derive from macroscopic potentials, through electrochemical effects (e.g. concentration gradients), electrokinetics (e.g. electrical potential gradients, electro-osmosis) or thermo-electrical effects (e.g. temperature gradients). Considering some of these potential fields, the total electrical current density can be written as:
\begin{equation}
{\bf{J}} = -{\cal{L}}_T \dfrac{\boldsymbol{\nabla} T}{T} + \boldsymbol{\nabla}\Phi_J + {\cal{L}}_{e}\boldsymbol{\nabla} V - {\cal{L}}_{ek}\boldsymbol{\nabla} P,
\label{eqn:Jtot}
\end{equation}
where $T$ is the temperature [K], ${\cal{L}}_T = -T\pi\sigma_T$ [W.m.s$^{-1}$] is given by the Peltier effect, $\Phi_J$ is the junction potential [V], which also can be expressed as a function of chemical concentration gradients in electrolytes ($\boldsymbol{\nabla}C_c / C_c$) by $\boldsymbol{\nabla}\Phi_J = \alpha_m\boldsymbol{\nabla}C_c / C_c$, with $\alpha_m$ the fluid junction coupling coefficient \citep{naudet03,main08,jouniaux09}. The parameter $V$ is the electrical potential [V], thus Ohm's law identify ${\cal{L}}_{e}= - \sigma_r$, with $\sigma_r$ the bulk electrical conductivity [S.m$^{-1}$]. The last term of equation \ref{eqn:Jtot} describes electrokinetic effects created by the driving pressure gradient $\boldsymbol{\nabla} P$, through the electrokinetic coupling ${\cal{L}}_{ek}$, defined as ${\cal{L}}_{ek} \equiv - \sigma_r C$ [A.Pa$^{-1}$.m$^{-1}$] \citep{pride94}. The parameter $C$ [V.Pa$^{-1}$] is known as the SP coefficient. The total water pressure P can be inferred from water pressures $h$ with: $P = \rho_wg(h - z)$, where $\rho_w$ is the water density [kg.m$^{-3}$], g is gravity and $z$ is the vertical location taken to be positive downward.\\
Considering a constant temperature, and no concentration gradients, one can write the following coupled equation:
\begin{eqnarray}
{\bf{J}} = {\cal{L}}_{e}{\boldsymbol{\nabla}}V - {\cal{L}}_{ek}{\boldsymbol{\nabla}}P,\\
\text{or},\nonumber \\
{\bf{J}} = -\sigma_r{\boldsymbol{\nabla}}V + \sigma_r C{\boldsymbol{\nabla}}P.
\label{eqn:J}
\end{eqnarray}
Without any external current sources, the conservation of the total current density implies,
\begin{equation}
{\boldsymbol{\nabla}}\cdot\mathbf{J} = 0.
\label{eqn:divJ}
\end{equation}
In the case of heterogeneous medium, one can assume for example a tabular medium with electrical conductivity and SP coefficient contrasts, then equation \ref{eqn:J} through \ref{eqn:divJ} leads to the following Poisson's equation:
\begin{equation}
\boldsymbol{\nabla}\cdot\mathbf{J} = -\sigma_r\boldsymbol{\nabla}^2V - \boldsymbol{\nabla}V\cdot\boldsymbol{\nabla}\sigma_r + \boldsymbol{\nabla}(C\sigma_r)\cdot\boldsymbol{\nabla}P + \sigma_rC\boldsymbol{\nabla}^2P = 0.
\label{eqn:poisson}
\end{equation}
In addition to primary current sources occuring from the term $\sigma_rC\boldsymbol{\nabla}^2P$, some secondary sources linked to the term $\boldsymbol{\nabla}(C\sigma_r)\cdot\boldsymbol{\nabla}P$ appear. These sources are located at boundaries formed by electrical conductivity and SP coefficient constrasts.\\
In the case of an homogeneous medium and without any contrasts of $\sigma_r$ and $C$, the equation \ref{eqn:poisson} reduces to:
\begin{equation}
\boldsymbol{\nabla}^2V = C\boldsymbol{\nabla}^2P.
\label{eqn:poisson_hom}
\end{equation}
Considering for example a steady-state saturated flow through a capillary the SP coefficient $C$ [V.Pa$^{-1}$] can be expressed as the ratio between the SP difference $\Delta V$ [V] and the driving-pressure difference $\Delta P$ [Pa],
\begin{equation}
C = \frac{\Delta V}{\Delta P}.
\label{eqn:Cdip}
\end{equation}
In one dimension, the equation \ref{eqn:divJ} can be written as,
\begin{equation}
\frac{\partial}{\partial z} \left(-\sigma_r \frac{\partial V}{\partial z} \right) + \frac{\partial}{\partial z} \left[{\rho_wg \cal{L}}_{ek} \left(\dfrac{\partial h}{\partial z} - 1 \right) \right] = 0.
\label{eqn:poisson_glob}
\end{equation}
\subsection{Discretization of the Poisson's equation}
\noindent
The Poisson's equation was solved using 1D finite element method. A first order basis function $\phi(z)$ has been chosen to discretize the Poisson's equation. This choice implies that variables and coefficients of the equation \ref{eqn:poisson_glob} vary linearly in each element. For each node $i$ of the system, the basis function is written $\phi_i(z)$. Then, the equation which has to be solved can be written as,
\begin{eqnarray}
\int_{0}^{l} \frac{\partial}{\partial z} \left(-\sigma_r \frac{\partial V}{\partial z} \right)\phi(z)dz \nonumber \\
+ \int_{0}^{l} \frac{\partial}{\partial z} \left[\rho_wg {\cal{L}}_{ek} \left(\dfrac{\partial h}{\partial z} - 1 \right) \right]\phi(z)dz = 0.
\label{eqn:poisson_galerkin}
\end{eqnarray}
Using this method, all variables and coefficients are approximated on each element using the same basis function:
\begin{eqnarray}
V(z,t) = \sum_{1}^{ne} V_i(t) \phi_i(z),\\
h(z,t) = \sum_{1}^{ne} {h}_i(t) \phi_i(z),\\
{\cal{L}}_{ek}(z,t) = \sum_{1}^{ne} {{\cal{L}}_{ek}}_i(t) \phi_i(z),\\
\sigma_r(z,t) = \sum_{1}^{ne} {\sigma_r}_i(t) \phi_i(z),
\end{eqnarray}
where $nn=ne+1$ is the number of nodes in the system of length $l$, and $ne$ the number of elements. The $V_i(t)$, ${h}_i(t)$, ${{\cal{L}}_{ek}}_i(t)$ and ${\sigma_r}_i(t)$ are respectively the values of the electrical potential, water pressure, electrokinetic coupling and bulk electrical conductivity at node $i$. For notations simplicity ${{\cal{L}}_{ek}}_i(t)$ will be written ${\cal{L}}_i(t)$ in the following. After integrating by parts equation \ref{eqn:poisson_galerkin}, the final system of equation to be solved is given by,
\begin{equation}
[A].V_i = - [B].h_i - \{F_j\},
\label{eqn:mat_syst}
\end{equation}
where the elements of [A], [B] and vector $\{F_j\}$ can be deduced from the following integrals:
\begin{eqnarray}
A_{ij} = \int_{0}^{l} \left( \sum_{k=1}^{nn} \phi_k\ \sigma_k \right)\phi^{'}_i\phi^{'}_j dz,\\
B_{ij} = \rho_wg \int_{0}^{l} \left( \sum_{k=1}^{nn} \phi_k\ {\cal{L}}_k \right)\phi^{'}_i\phi^{'}_j dz,\\
F_{j} = \rho_wg \int_{0}^{l} \left( \sum_{k=1}^{nn} \phi_k\ {\cal{L}}_k \right)\phi^{'}_j dz.
\end{eqnarray}
The matrices [A] is tridiagonal and the system can be solved using Thomas algorithm \citep{press92}. The system introduced in equation \ref{eqn:mat_syst} can be detailed for each element $i$ as:
\begin{eqnarray}
\dfrac{1}{2\Delta z} \left[ - (\sigma_{i-1}+\sigma_i).V_{i-1} + (\sigma_{i-1}+2\sigma_i+\sigma_{i+1}).V_i \right. \nonumber \\
\left.- (\sigma_{i}+\sigma_{i+1}).V_{i+1} \right] = \dfrac{\rho_wg}{2\Delta z} \left[ ({\cal{L}}_{i-1} + 2{\cal{L}}_{i}) + {\cal{L}}_{i+1}).h_i  \right. \nonumber \\
\left. - ({\cal{L}}_{i-1} + {\cal{L}}_{i}).h_{i-1} - ({\cal{L}}_{i} + {\cal{L}}_{i+1}).h_{i+1} + ({\cal{L}}_{i-1} + {\cal{L}}_{i+1})  \right]
\end{eqnarray}
where $\Delta z = 0.1$ cm is the spatial discretization. Boundary conditions at nodes $1$ and $nn$ are:
\begin{eqnarray}
\sigma_1 \dfrac{\partial V}{\partial z} \phi_1 = - j_0 \\
\sigma_{nn} \dfrac{\partial V}{\partial z} \phi_{nn} = -j_l
\end{eqnarray}
with $j_0$ and $j_l$ are the values of the current density at the top and the bottom of the system respectively. The Poisson's equation is solved in terms of electrical potential $V$. Thus, two boundary conditions are necessary: one on the total current density $J$ (Neumann type) and one on the electrical potential itself $V$ (Dirichlet type). As no current outflow can occur at the top and the bottom of the system (i.e no electrical exchange between the medium and the air), the total current density $j_0$ at $z=0$ and $z=l$ are: $j_0 = j_l = 0$ A.m$^{-2}$. In addition, a constant value of electrical potential $V_0$ has been chosen at the top of the system. This constant can be a reference value, so that $V_0 = 0$ V has been chosen for simplicity.\\
The modelling process is carried out using the following protocol: 1. Considering hydrodynamic boundary conditions, the equation \ref{eqn:richards} is first solved, so that water pressures and corresponding water contents are computed at each node and each time step; 2. The Poisson's equation is then solved using equation \ref{eqn:mat_syst}, which allows to compute the electrical potential at each node of the system. 3. Thus, the needed potential differences can be deduced from the computed electrical potential field and compared to measurements. The next section presents a test performed considering a linear water pressure profile in saturated conditions, i.e simulating a Darcy's experiment. The following sections go into the unsaturated case and present computed SP differences and measurements in the case of a drainage experiment. Several hypotheses on SP coefficient dependence on water content will be considered and compared.
\section{Saturated flow modelling}
\begin{figure*}
\begin{center}
\includegraphics[width=16cm]{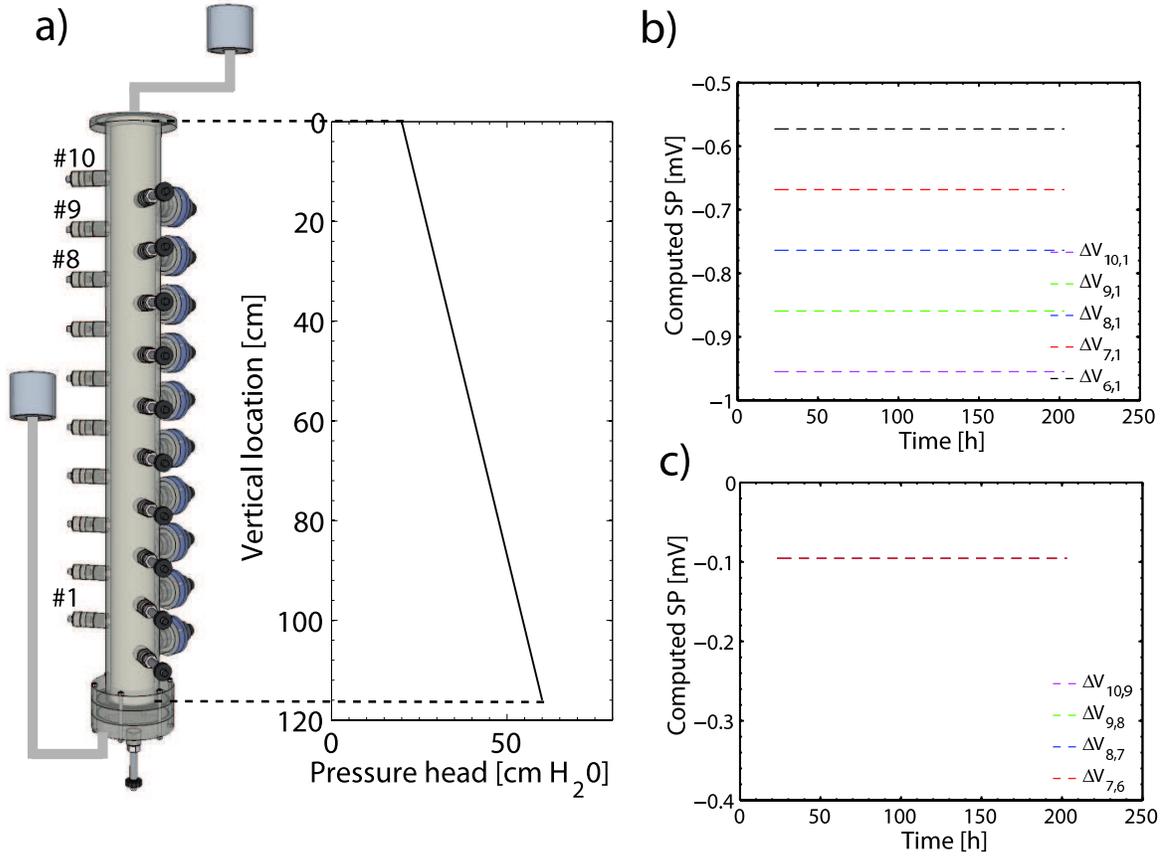}
\end{center}
\caption{a) The water pressure profile used for the Darcy's experiment simulation. The pressure head boundary conditions are constant and equal to $h_{0}=20$ cm and $h_{l}=60$ cm of equivalent water height at the top and the bottom of the column respectively. b) Computed SP between the first five electrodes (\#10 to \#6) and the reference \#1. c) Computed SP for the dipoles (10,9), (9,8), (8,7) and (7,6). The computed values are $\Delta V{i,i-1} = -0.0954$ mV and $\Delta P{i,i-1} = 59.67$ Pa which lead to $C_{sat} = 1.6\times$10$^{-6}$ V.Pa$^{-1}$.}
\label{fig:profil_h}
\end{figure*}

\noindent
The first step is to test the code accuracy by modelling a saturated flow through a column of sand ($l=1.16$ meter height), i.e a Darcy's experiment. The complete scheme described above was applied, so that equations \ref{eqn:richards} and \ref{eqn:poisson} were solved. The water electrical conductivity $\sigma_w$ and saturated SP coefficient $C_{sat}$ corresponding to this test are those of \citet{allegre2010} and are reported in Table \ref{tab:par_hom}. A linear and steady-state water pressure profile was applied to the medium (Figure \ref{fig:profil_h}a). Thus, boundary conditions at the top and the bottom of the column are constant water pressure head and given by: $h_0(t) = 20$ cm and $h_l(t) = 60$ cm. In this case the medium remains saturated, so that $S_w = 1$. The boundary conditions for Poisson's equation was a zero current density $J_0$ at the top and a constant electrical potential $V_0=$0 V at the bottom of the column. The first simulated electrode (\#10) is located at 11 cm from the top of the system, while other electrodes are placed each 10 cm along it. The SP are then computed between the first five electrodes (\#10 to \#6) and the reference electrode (\#1 15 cm above column's bottom) and for each dipole formed by two consecutive electrodes (e.g $\Delta V_{10,9}$ will be the computed electrical potential difference between electrodes \#10 and \#9). The simulation were performed for 200 hrs, and the same results were obtained at each time step.\\
The SP differences computed between each electrode and the reference are different. This is coherent with the increasing size of each dipole, so that larger the dipole is, larger the SP difference is (Figure \ref{fig:profil_h}b). As the medium is homogeneous (fully saturated) SP differences computed for each dipole are merged (Figure \ref{fig:profil_h}c) and exhibit the value $\Delta V_{i,i-1} = - 0.0954$ mV. The corresponding driving pore pressure difference for each dipole is $\Delta P_{i,i-1} = 59.67$ Pa. Considering these values, the inferred SP coefficient at saturation computed with equation \ref{eqn:Cdip} is: $C_{sat} = -$ 1.6$\times$10$^{-6}$ V.Pa$^{-1}$. This value is exactly equal to the original $C_{sat}$ (see Table \ref{tab:par_hom}) used for calculation. This test experiment was performed using others water pressure profiles, other $C_{sat}$ values and other boundary condition value $V_0$ and always yielded to the original $C_{sat}$. We therefore confirm that the apparent measurement of $\Delta V / \Delta P$ for the dipoles can provide the streaming potential coefficient for saturated conditions.
\section{Drainage experiment}
\subsection{Unsaturated flow modelling}
\begin{figure}
\begin{center}
\includegraphics[width=8cm]{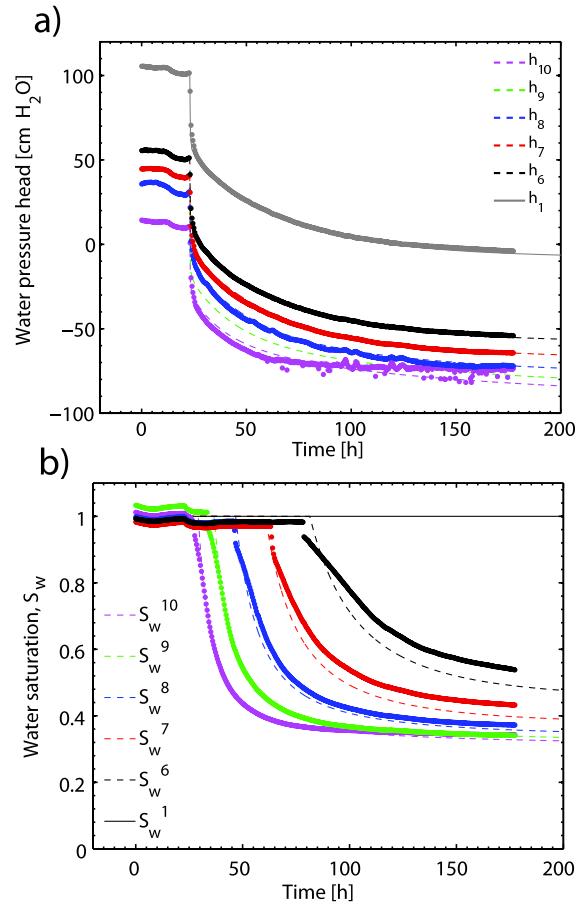}
\end{center}
\caption{a) Measured (dots) and computed (dashed lines) water pressure heads deduced from the Richards equation solving. Indices $i$ indicates the location of measurement, as $h_{10}$ is the pressure head 16 cm from the column top and $h_{1}$ 10 cm from the column bottom. Other $h$ values are measured and computed each 10 cm. b) Measured (dots) and computed (dashed lines) water saturations deduced from the Richards equation solving, where indices $i$ indicates the location of measurement. The drainage starts at $t\simeq 22$ hr.}
\label{fig:h_theta}
\end{figure}

\noindent
We propose to model the SP measurements of a drainage experiment carried out by \citet{allegre2010}. The SP were measured during a drainage experiment performed in a column of plexiglass of approximately 1.2 metre height and 10 cm diameter fullfilled with clean Fontainebleau sand. A constant water pressure applied at the bottom of the column allowed the drainage to start. \citet{allegre2010} combined SP measurements to water content and water pressure measurements each 10 cm along the column. The aim of this section is to computed SP using the numerical scheme described above, considering several hypotheses for the SP coefficient dependence on water content. Thus, formula for $C_r$ of \citet{guichet03}, \citet{revil07} and a new empirical model will be implemented to solve the Poisson's equation.\\
At the beginning of the experiment, the medium is fully saturated, so that the water pressure profile is hydrostatic. The drainage starts when the boundary condition at the bottom of the column is set to a value of $h=2$ cm of equivalent water height (i.e $p_w(z=l)\simeq$ 200 Pa). The water pressure heads and water contents are computed at each time step by solving the equation \ref{eqn:richards} and are compared to experimental data (Figure \ref{fig:h_theta}). The parameters of the retention model (eq. \ref{eqn:brookscorey}) and the relative permeability model (eq. \ref{eqn:mualem}) are reported in Table \ref{tab:hydro}. Since pressure sensors are located each ten centimetre along the column, before the drainage start, the pressure heads measurements are shifted from 10 cm between each other. These measurements are characteristics for the hydrostatic equilibrium. After the drainage start, the measured pressure heads decrease all at the same time and stabilize at different values depending on the sensor location. This shift of the pressure values at the end of the experiment (e.g at t$\simeq$ 150 h) indicates the presence of a capillary fringe and of a water saturation gradient (see Figure \ref{fig:h_theta}b). For very long time (around 90 days for the sand used) the water phase equilibrates itself in the column until a linear water pressure profile is reached. From the drainage start, the water saturations do not decrease at the same time, but one after the other depending on the measurement location. The time shift between the water content measurements informs on the dynamic of the saturation front propagation during the drainage experiment.\\
The hydrodynamic parameters used in the Richards' equation (Table \ref{tab:hydro}) were computed by inversion of the water pressure heads and water content measurements. The detailed inverse problem procedure can be found in \citet{hayek08}. It is shown that this approach provides good estimations and small errors on $\theta_s$, $\lambda$ and $h_a$.
\subsection{Unsaturated SP modelling}
\noindent
To solve the Poisson's equation at each time step, a model for the SP coefficient dependence on water saturation is needed. \citet{guichet03} proposed that the SP coefficient vary linearly with the effective water saturation,
\begin{equation}
C_r = S_e.
\label{eqn:mod_gui}
\end{equation}
\citet{revil07} had a different approach and proposed another relation depending on a relative permeability model as,
\begin{equation}
C_r = \dfrac{k_r}{S_w^{n+1}}~~\text{with}~~k_r = S_e^{L+2+2/\lambda},
\label{eqn:mod_rev}
\end{equation}
where $n$ is the Archie's saturation exponent \citep{archie42}. The $L$ parameter is usually chosen as $L=$0.5 \citep{mual76}, but is equal to 1 in \citet{revil07}, so that the two cases will be tested in the following. These two models have in common to predict the maximum of the SP coefficient to be $C_{sat}$ (for $S_w=1$). Moreover, they imply a monotonous decrease of the relative SP coefficient with decreasing saturation. \citet{allegre2010} recently observed that $C_r$ could be until 200 times larger than $C_{sat}$, and that it first increases with decreasing saturation, and then decreases up to the residual water saturation. So that, we propose an empirical relation for the SP coefficient inferred from these measurements written as,
\begin{equation}
C = C_{sat} S_e[1+\beta(1 - S_e)^\gamma]
\label{eqn:mod_nonmono}
\end{equation}
where $\beta$ and $\gamma$ are two fitted parameters. Note that $\beta$ depends on the considered dipole and varies as a function of the vertical location. This assumption coming from the experimental SP coefficients, which exhibit different maximum values, will be discussed in the following section. Contrary to the two first relations, this model predicts a non-monotonous behaviour of the SP coefficient as a function of water saturation. The three presented models were used to compute $C(S_e)$ from computed water saturations, to solve the Poisson's equation.\\
Moreover, an {\it{a priori}} is needed for the electrical conductivity $\sigma_r(S_w)$ to solve equation \ref{eqn:poisson}. The electrical conductivity is inferred at each time step using the Archie's law \citep{archie42},
\begin{equation}
\sigma_r = \sigma_w\phi^{m}S_w^{n}
\label{eqn:archie}
\end{equation}
with $m$ and $n$ the two Archie's exponent, $\sigma_w$ the water electrical conductivity [S.m$^{-1}$] and $\phi$ the porosity. The relation $C(S_e)$ is also computed at each time step and for each element of the system from computed water saturations. All parameters' values needed for computation are reported in Tables \ref{tab:hydro} and \ref{tab:par_ek}. Using the computed electrical potential values in the column, SP were computed for the dipoles (10,9), (9,8), (8,7) and (7,6), which are the dipoles located in the unsaturated part of the column during the drainage experiment.
\section{Discussion}
\begin{figure*}
\begin{center}
\includegraphics[width=\textwidth]{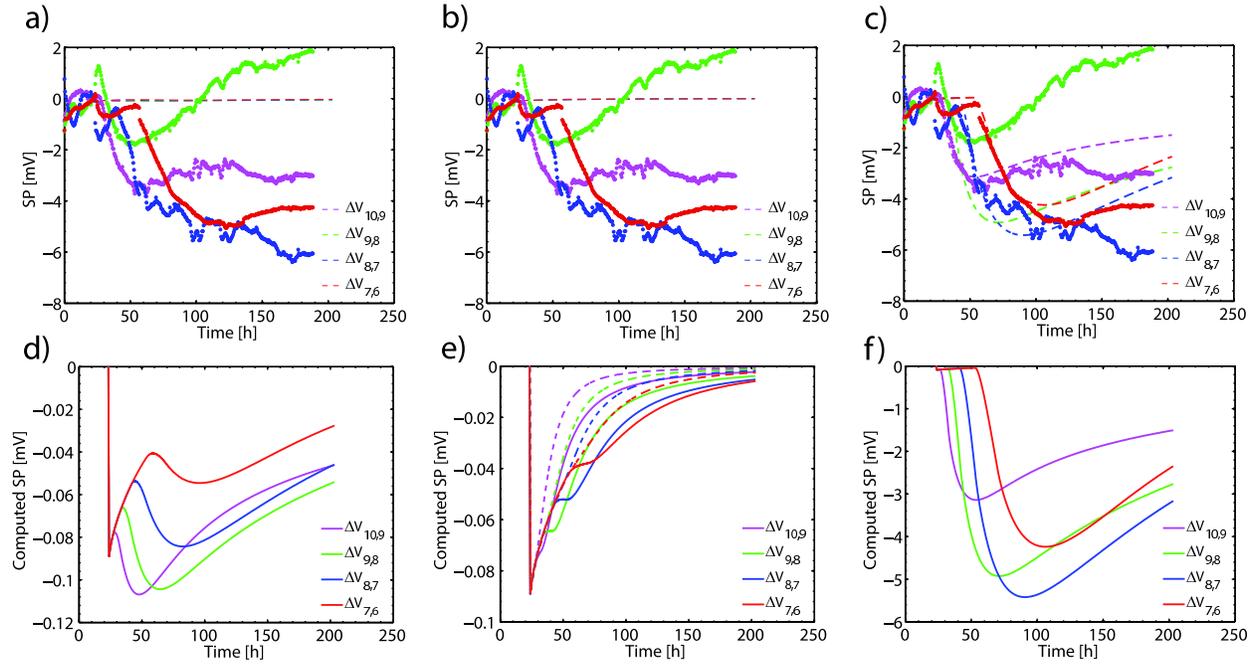}
\end{center}
\caption{Measured (dots) and computed (lines) streaming potentials deduced from the Poisson's equation solving, for each dipole, using respectively equations \ref{eqn:mod_gui} (a), \ref{eqn:mod_rev} (b) and \ref{eqn:mod_nonmono} (c). Computed SP using equations \ref{eqn:mod_gui} (d), \ref{eqn:mod_rev} (e) and \ref{eqn:mod_nonmono} (f) in the Poisson's equation, and parameters from tables \ref{tab:hydro} and \ref{tab:par_ek}. The \citet{revil07} model (e) has been implemented using $L=$0.5 (lines) and $L=$1 (dashed lines).}
\label{fig:SP_glob}
\end{figure*}

\begin{figure}
\begin{center}
\includegraphics[width=8cm]{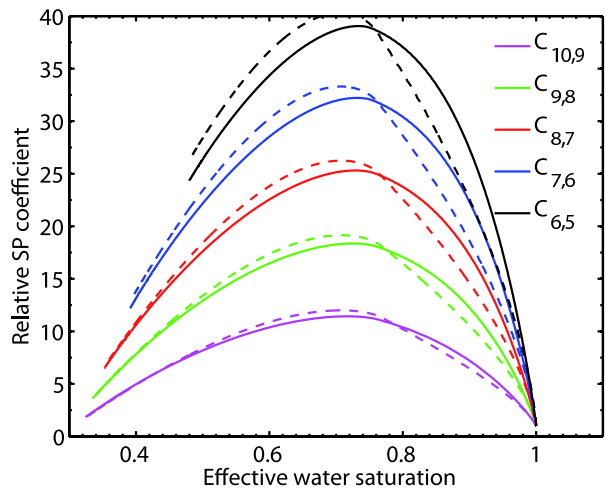}
\end{center}
\caption{Computed relative SP coefficients using eq. (\ref{eqn:Cdip}) with computed $\Delta V$ and $\Delta P$ after Poisson's equation was solved (lines), and using eq. (\ref{eqn:mod_nonmono}) (dashed lines), for five locations in the column.}
\label{fig:CSe_calc}
\end{figure}

\begin{figure}
\begin{center}
\includegraphics[width=8cm]{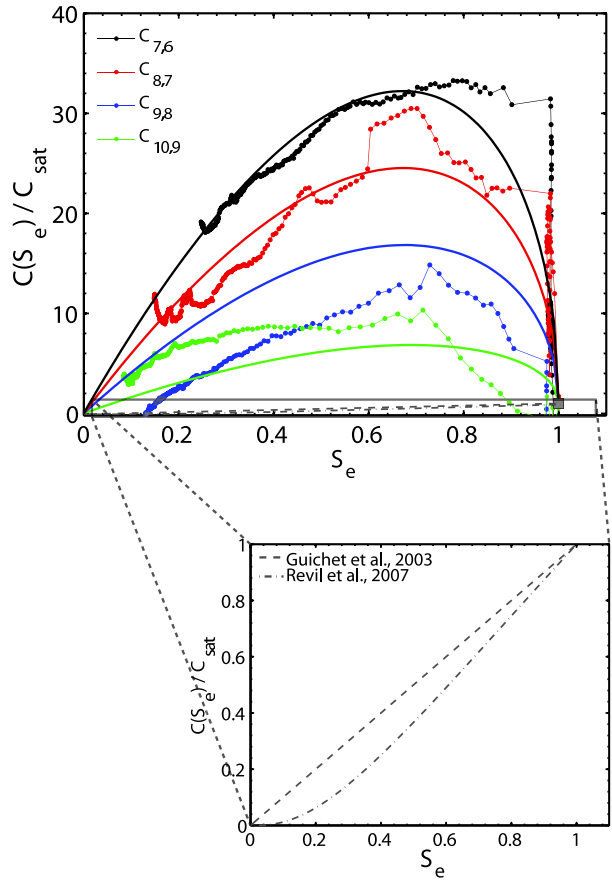}
\end{center}
\caption{(On the top) Experimental SP coefficients $\Delta$V / $\Delta$P from \citet{allegre2010} (dots) and model adjusted to measurements using equation \ref{eqn:mod_nonmono}). The $\beta$ and $\gamma$ values are reported in the Table \ref{tab:par_ek}. (On the bottom) The \citet{guichet03} and \citet{revil07} models for the relative SP coefficient $C_r$ as a function of water saturation.}
\label{fig:SP_coeff_mod}
\end{figure}

\begin{figure*}
\begin{center}
\includegraphics[width=\textwidth]{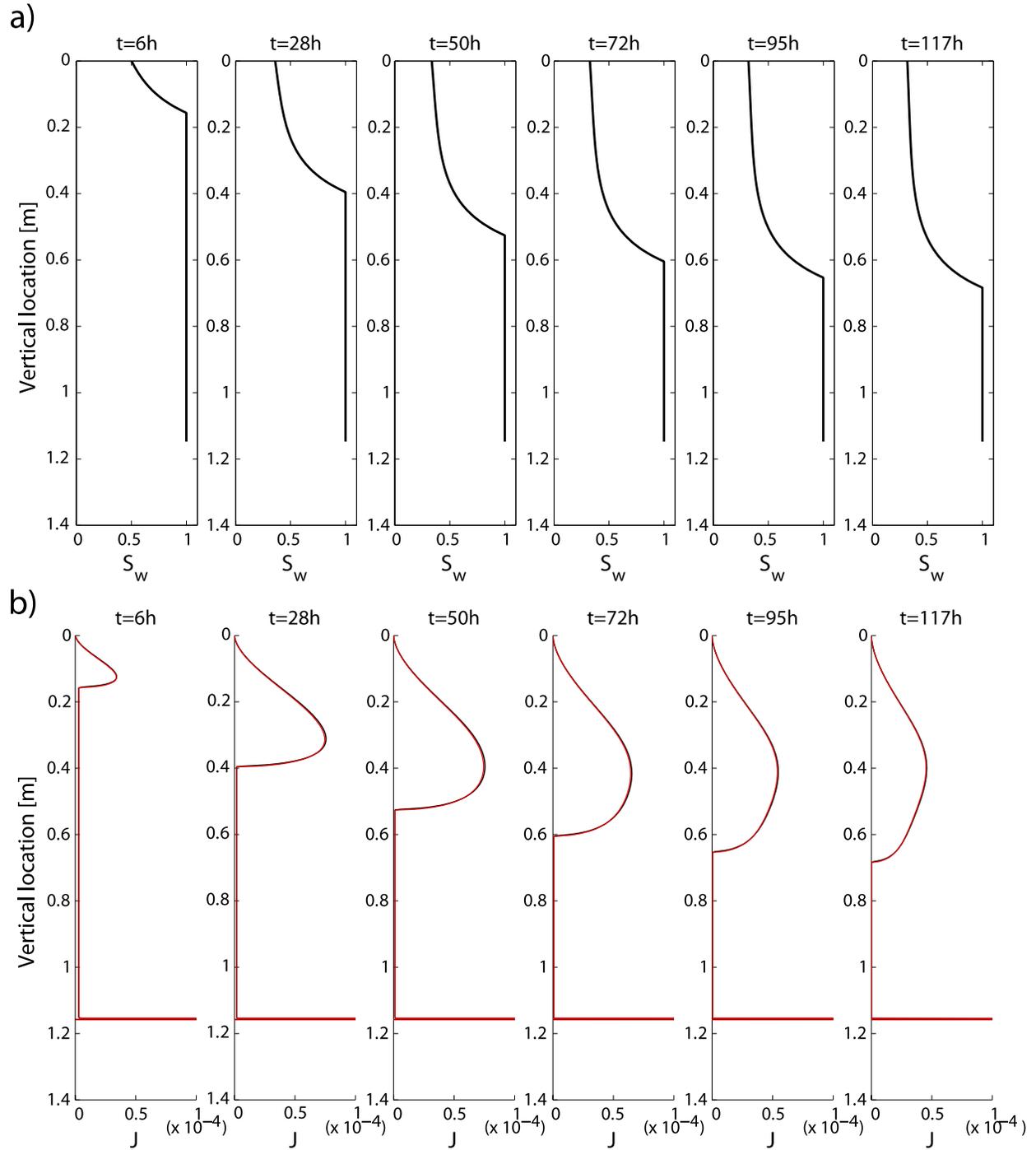}
\end{center}
\caption{Snapshots of the water saturation profile (a) and total current density components $J_{cond}$ (black line) and $J_{conv}$ (red line, b) for six times between $t=6$ h and $t=117$ h. The electrical current density is expressed in A.m$^{-2}$. The snapshot corresponds to the simulation performed using the equation \ref{eqn:mod_nonmono} for implementation of $C(S_e)$ in the Poisson's equation.}
\label{fig:Jtot}
\end{figure*}

\begin{figure}
\begin{center}
\includegraphics[width=8cm]{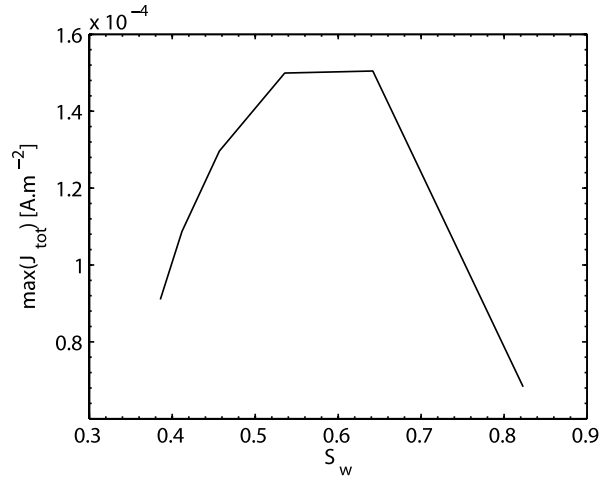}
\end{center}
\caption{Maximum of the total current density $J$ (black line) inferred from curves in Figure \ref{fig:Jtot} and corresponding water flow velocity (dashed black line).}
\label{fig:maxJ}
\end{figure}

\begin{figure*}
\begin{center}
\includegraphics[width=\textwidth]{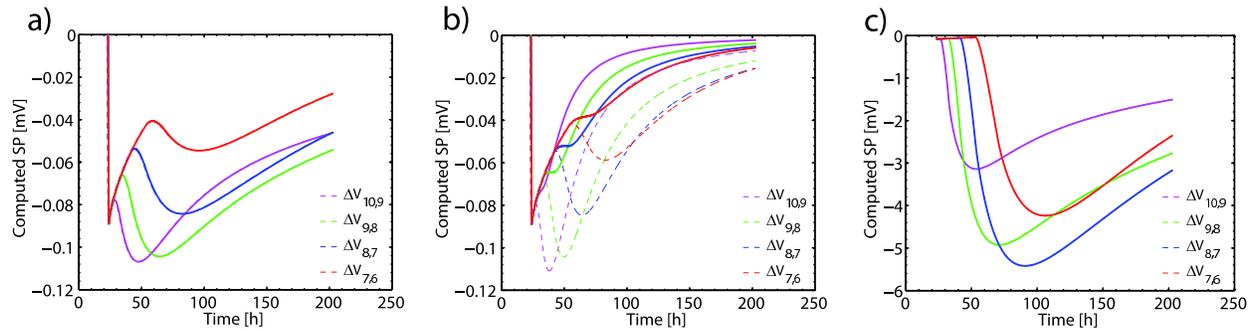}
\end{center}
\caption{Computed SP (lines) using equations \ref{eqn:mod_gui} (a), \ref{eqn:mod_rev} (b) and \ref{eqn:mod_nonmono} (c) with $n = 1.45$, compared to computed SP with $n = 2.5$ (dashed lines).}
\label{fig:SP_sig}
\end{figure*}

\noindent
The SP were computed solving equations \ref{eqn:richards} and \ref{eqn:poisson} for the three presented models (Figure \ref{fig:SP_glob}). The computed SP using \citet{guichet03} and \citet{revil07} models are very small compared to the measurements (Figure \ref{fig:SP_glob}a,b). A jump, corresponding to the drainage start, is observed in SP signals at the beginning of the simulation. Its magnitude is around 0.09 mV for tests performed using equations \ref{eqn:mod_gui} and \ref{eqn:mod_rev} and is the same for all dipoles (Figure \ref{fig:SP_glob}d,e). In the case of \citet{guichet03} model (Figure \ref{fig:SP_glob}d), the computed SP begin to decrease to reach a minimum value for dipoles (10,9) and (9,8) around -0.11 mV and then increase. This minimum is observed at the time when water saturation stops to decrease for all dipoles. In the case of \citet{revil07} model, the increasing of computed SP is monotonous during the simulation, from a value around -0.09 mV to almost 0 for all dipoles. It is obvious that these two models are not appropriate to explain the SP measurements both in terms of magnitude and behaviour.\\
\citet{revil07} used $L=$1 in their model, instead of $L=$0.5 as proposed by \citet{mual76} to hold as the best value for forty-five soils including sand. Consequently, the equation \ref{eqn:mod_rev} was also implemented using $L=$0.5, which leads to a modified \citet{revil07} model. The results (Fig. \ref{fig:SP_glob}e) are similar to those for $L=$1 in terms of amplitude, but show an increasing of SP signals without any step as it was observed before. The choice of $L$ is quite important because it's involved in the global power law in eq. \ref{eqn:mod_rev}, and consequently influences the shape of the model \citep{allegre2010}.\\
On the other hand, the model introduced for $C_r$ by equation \ref{eqn:mod_nonmono} leads to good results in terms of computed SP. The measured SP signals are well reproduced, particularly at the beginning of the experiment between $t\simeq 20$ h and $t\simeq 100$ h. The computed SP corresponding to dipoles (10,9) to (7,6) decrease one after the other when water saturation (measured at the same level) begins to decrease (see Fig. \ref{fig:h_theta}b and \ref{fig:SP_glob}c). Thus, the saturation front propagation is characterized by the time shift between SP decreasing starts. The computed SP using both equations \ref{eqn:mod_gui} and \ref{eqn:mod_rev} are up to one and a half order of magnitude smaller than those computed using expression \ref{eqn:mod_nonmono} (Fig. \ref{fig:SP_glob}).\\
For times over 100 hours, the residuals between measured and computed SP are larger. At this point of the experiment the water flow is very low, leading to very small variations of SP. This point is interpreted in \citet{allegre2010}, who precise that for such low water outflow (even not measurable), the SP variations are too weak to insure robust interpretation. Consequently, a better fit was expected at the beginning of the experiment (when the water flow is maximum), which is the case for three measurement dipoles. In addition, all the tests performed to ensure the quality of SP recordings and prevent the external sources of noise from disturbing the measurements, including a statistical study on uncertainties, can be found in \citet{allegre2010} (Appendices A, B). Moreover, the fit and estimation of the set of parameters of eq. \ref{eqn:mod_nonmono} would be improved if the SP measurements were inverted. Thus a joint inversion would give more informations on parameters sensitivities.\\
One can verify the assumption made by \citet{allegre2010} to use eq. (\ref{eqn:Cdip}) to inferred SP coefficients from SP measurements. Thus, SP coefficients was computed using eq. (\ref{eqn:Cdip}), and compared to SP coefficients computed with eq. (\ref{eqn:mod_nonmono}) averaged on 10 cm to be representative of the same investigated volume (Fig. \ref{fig:CSe_calc}). It is shown that SP coefficients using eqs (\ref{eqn:Cdip}) and (\ref{eqn:mod_nonmono}) give very close results in terms of behaviour and magnitude. This example prove the validity of using eq. (\ref{eqn:Cdip}) to infer accurate values of $C$ even for non-steady conditions. One can consider that these values are apparent SP coefficients because the water saturation distribution is not perfectly homogeneous in 10 cm of sand, however they are still perfectly representative of true SP coefficients, at least for this kind of drainage experiment.\\
Furthermore, computed SP coefficients using different assumptions for $C_r$ (equations \ref{eqn:mod_gui}, \ref{eqn:mod_rev} and \ref{eqn:mod_nonmono}) can be compared to measurements of $\Delta V$/ $\Delta P$ from \citet{allegre2010} (Fig. \ref{fig:SP_coeff_mod}). The fitted values of $\beta$, $\gamma$, and parameters of equations \ref{eqn:mod_gui} and \ref{eqn:mod_rev} are reported in the Table \ref{tab:par_ek}. It is clear on figure \ref{fig:SP_coeff_mod} that the two existing models for $C_r$ predict lower values of $C_r$ than measurements and fail to provide the correct behaviour of $C_r$. Moreover, the measured SP coefficients $\Delta V$/ $\Delta P$ are quite well reproduced in the case of using equation \ref{eqn:mod_nonmono}. It is important to notice that \citet{allegre2010} inferred experimental SP coefficients using the equation \ref{eqn:Cdip} which is classically used for steady-state flows, neglecting electrokinetic sources coming from SP coefficient and electrical conductivity contrasts. Since no electrokinetic sources have been neglected in the present modelling, one can conclude that the non-monotonous behaviour combined to large values of $C_r$ observed previously are not artefacts and are not created by neglected contrasts of $C_r$ or $\sigma_r$ but has a physical origin. Therefore we show that the measurement of apparent $\Delta V / \Delta P$ for the dipoles can provide the streaming potential coefficient $C_r$ in these conditions. Nevertheless, a slight difference is observed between the maximum of computed and measured SP coefficients. These differences come from the use of the equation \ref{eqn:Cdip} to infer SP coefficients. Finally, these results confirm that a different maximum in $C_r$ for each dipole is necessary to provide accurate computed SP differences. We suggest that this observation has a physical meaning coming from the flow behaviour. Indeed, each dipole is not affected by the same hydrodynamic conditions, and particularly the same flow velocity, when water saturation (at its level) decreases.\\ 
An important point to discuss is the behaviour of the total current density. The figure \ref{fig:Jtot} shows six snapshots of both water saturation and current density component $J_{cond} = -\sigma_r \boldsymbol{\nabla} V$ and $J_{conv} = \sigma_rC \boldsymbol{\nabla} P$ vertical profiles at six time during the simulation. The components $J_{cond}$ and $J_{conv}$ are computed {\it{a posteriori}} using finite differences. Even if SP sources are linked to $\boldsymbol{\nabla} \cdot$ {\bf{J}}, as introduced by the conservation equation (eq. \ref{eqn:divJ}), we think that the vertical component of {\bf{J}} is still relevant to describe the general behaviour of SP signals. This representation is usefull {\it{a posteriori}} to interpret the global SP variations, as it integrates all electrokinetic sources coming from SP coefficient or electrical conductivity contrasts (see eq. 10). The water saturation vertical profiles characterizes the propagation of saturation front during the drainage. In the saturated part of the column, $\theta = \theta_s$ thus $S_w =1$. In the unsaturated part, the saturation decreases from the top to the saturation front. The saturation value at the top of the system decreases during drainage from $S_w(z=0) = 0.5$ at $t=6$ h to $S_w(z=0) = 0.33$ at $t=50$ h and then stabilizes. For times greater than $t=50$ h, the water saturation is almost constant ($S_w \simeq 0.33$) for $0 < z < 40$ cm. Moreover, it is shown on figure \ref{fig:Jtot}b that the conduction ($J_{cond}$) and the convection ($J_{conv}$) component of the total current density $J$ are almost equal in absolute terms in the whole system during the entire experiment. The maximum difference observed between computed $J_{cond}$ and $J_{conv}$ does not exceed 2 per cent. This is a crucial point because the relation $J_{cond}=J_{conv}$ allows the use of equation \ref{eqn:Cdip} (i.e $C = \Delta V / \Delta P$) to deduce SP coefficient values from measurements. This is another confirmation that this approach can be used and gives accurate results for this kind of experimental conditions.\\
A large discontinuity is observed at the bottom of the column. This comes from the important contrast of electrical conductivity between the saturated medium and the water in the reservoir at the bottom of the column. Thus, the electrical conductivity changes from $\sigma_{sat} \simeq 0.002 $ S.m$^{-1}$ to $\sigma_w = 0.01 $ S.m$^{-1}$ at this boundary. This is the only possible source of current since the SP coefficient is zero in water. Nevertheless, this contrast does not influence the SP measurements, since the measurement are located far from it and that the current density returns to a constant value in the centimetre above it. It was suggested by \citet{linde07} and Revil and Linde ({\it{comment submitted}}), that this contrast could be responsible for the behaviour of the measured SP. However, it is not the case for our experiment but may occur for measurements involving an electrode very close to this source. It should be then considered as an experimental artefact in this case.\\
The second point is that the maxima of $J_{cond}$ and $J_{conv}$ are not located at the higher gradient of water saturation (at the saturation front), corresponding to the higher contrast of electrical conductivity $\sigma_r$, but backward from this front. Its absolute value varies not linearly from $3\times$ 10$^{-5}$ A.m$^{-2}$ to $6.8\times$ 10$^{-5}$ A.m$^{-2}$ for 6 h $< t <$ 117 h (Figure \ref{fig:maxJ}). This suggests that the predominant contribution to the total electrical density comes from contrasts in SP coefficient which is larger for $S_w \simeq 0.8$ than for saturated conditions. This is obviously a consequence of the behaviour of $C_r$ as a function of water saturation. Thus, it seems that the electrokinetic response could be dominated by SP coefficient gradients and not by electrical conductivity gradients for such water flow.\\
This statement can be investigated using different parameters in the Archie's law (eq. \ref{eqn:archie}) to increase the influence of the electrical conductivity. Thus, the previous value of $n=1.45$ is replaced by a larger value as $n=2.5$. The influence of the saturation exponent $n$ depends on the considered model for the SP coefficient (Fig. \ref{fig:SP_sig}a,b,c). Indeed, any change in $n$ does not influence the computed SP in the case of using equations \ref{eqn:mod_gui} or \ref{eqn:mod_nonmono} for calculation. On the contrary, the increasing of $n$ from $1.45$ to $2.5$ changes both the behaviour and magnitude of computed SP when the equation \ref{eqn:mod_rev} is implemented (Fig. \ref{fig:SP_sig}b). Thus, resulting SP exhibit larger values than the previous ones. This is explained by the increasing of the SP coefficient value implied by the increasing of $n$ (see eq. \ref{eqn:mod_rev}). It suggests that electrical conductivity contrasts are insignificant on the computed SP compared to other electrokinetic sources induced by SP coefficient contrasts.\\
One can conclude that a non-monotonous behaviour combined to large values of $C(S_w)$ needs to be implemented in the Poisson's equation to obtain computed SP close from the measured one. Nevertheless, the expression presented in this work is not sufficient and our approach does not deal with physical considerations. 
In future work, the water-flow conditions of the drainage experiment and especially the pressure dynamic and flow velocity could be involved in the observed SP coefficient behaviour.    
\section{Conclusions}
\noindent
We presented in this work the first modelling of both Richards and Poisson's equations for unsaturated conditions using 1D finite element method. Several simulations have been performed using three different hypotheses on the SP coefficient $C_r$ to deduce SP differences which have been compared to observations from \citet{allegre2010}. The two existing models from \citet{guichet03} and \citet{revil07} were not able to predict SP differences consistent with the measurements of \citet{allegre2010}, although no electrical current sources have been neglected. We proved with this modelling approach that a non-monotonous expression for $C(S_e)$ is necessary to correctly reproduced these measurements. We also demonstrated the consistency of our SP coefficient dataset verifying the equality $J_{cond}=J_{conv}$, and consequently the possibility to use the apparent $\Delta$V / $\Delta$P measurements to infer correct $C$ values for this kind of drainage experiment. This conclusion has to be verified for other flow conditions, for example steady-state unsaturated flow conditions with various flow velocities using different sands.\\
Finally, a joint inversion approach of hydrodynamics and electrical potential, which could improve the modelling results of SP, is undergoing. This approach will help to insure the robustness of our model for long times and will define precisely the sensitivities of inverted model parameters.\\
For field applications, even if a good correlation can be observed between SP response and precipitation occurrence \citep{thony97,doussan02}, it seems that a linear relationship between water flux and electrical potential gradient would be difficult to establish. Indeed, non-linear effects coming from the behaviour of the SP coefficient for unsaturated conditions, show that this relationship is probably more complex. Moreover, some acquisition issues and changing soil conditions during long experiments make SP difficult to interpret \citep{doussan02}. However, shorter artificial infiltration experiments using SP monitoring could still be possible at the field scale. SP could be modelled with our new model of relative SP coefficient taking into account for infiltration and/or evaporation with time-varying upper boundary conditions. These experiments could be very useful to infer some hydrodynamic parameters of soils, such as hydraulic conductivity, and give a way to measure ground water flux in the vadose zone.

\section{Acknowledgments}
This work was supported by the French National Scientific Centre (CNRS), by ANR-TRANSEK, and by REALISE, the Alsace Region Research Network in Environmental Sciences in Engineering and the Alsace Region.

\bibliographystyle{gji}
\bibliography{biblio}

\newpage
\begin{table}
\caption{Parameters used to perform the Darcy's experiment simulation, deduced from \cite{allegre2010}.}
\begin{center}
\begin{tabular}{cccc}
\hline
$\sigma_w$ [S.m$^{-1}$] & $C_{sat}$ [V.Pa$^{-1}$] & $h_{0}$ [cm H$_2$O] & $h_{l}$ [cm H$_2$O] \\
\hline
103.2$\times$10$^{-4}$ & -1.6$\times$10$^{-6}$ & 20 & 60 \\
\hline
\end{tabular}
\end{center}
\label{tab:par_hom}
\end{table}
\begin{table*}
\caption{Hydrodynamic parameters values used to solve the Richards equation for unsaturated conditions. The parameter $K_s^{meas}$ is the measured permeability of the sand. These two values of permeability lead to the same results for pressure and water-content behaviours.}
\label{tab:hydro}
\begin{tabular}{@{}ccccccc}
\hline
$K_s $ (x$10^{-5}$) [m.s$^{-1}$] &$K_s^{meas} $ (x$10^{-5}$) [m.s$^{-1}$] & $h_a$ [m] & $\lambda$ & $\theta_r$ [-] & $\theta_s, \phi$ [-] & ${S_w}^r$ [-] \\
\hline
1.65 & 17.2 & 0.4 & 3.88 & 0.11 & 0.36 & 0.305\\
\hline
\end{tabular}
\end{table*}
\begin{table*}
\caption{Parameters needed to implement equations \ref{eqn:mod_rev}, \ref{eqn:mod_nonmono} and \ref{eqn:archie} in the Poisson's equation. The four values given for $\beta$ corresponds to the dipoles from (10,9) to (7,6). The $n$ value was measured in \citep{allegre2010}.}
\label{tab:par_ek}
\begin{tabular}{@{}ccccc}
\hline
$C_{sat}$ [V.Pa$^{-1}$] & $\sigma_w$ [S.m$^{-1}$] & $n$ (Archie exponent) & $\beta$ & $\gamma$ \\
\hline
-1.6$\times$10$^{-6}$ & 103.2$\times$10$^{-4}$ & 1.45 & 32,52,72,92 & 0.4 \\
\hline
\end{tabular}
\end{table*}
\newpage
\newpage
\newpage
\newpage
\newpage
\newpage
\newpage
\newpage

\end{document}